\documentclass[review]{elsarticle}

\usepackage{lineno,hyperref}
\usepackage{url}
\modulolinenumbers[5]

\journal{Journal of \LaTeX\ Templates}

\begin{document}

\begin{frontmatter}

\title{Repeating Pulsed Magnet System for Axion-like Particle Searches and Vacuum Birefringence Experiments}

\author[myaddress1]{T. Yamazaki\corref{mycorrespondingauthor}}
\cortext[mycorrespondingauthor]{Corresponding author. Tel.: +81 338158384}
\ead{yamazaki@icepp.s.u-tokyo.ac.jp}
\author[myaddress1]{T. Inada}
\author[myaddress1]{T. Namba}
\author[myaddress1]{S. Asai}
\author[myaddress1]{T. Kobayashi}
\author[myaddress2]{A. Matsuo}
\author[myaddress2]{K. Kindo}
\author[myaddress3]{H. Nojiri}

\address[myaddress1]{Department of Physics, Graduate School of Science, and International Center for Elementary Particle Physics, the University of Tokyo, 7-3-1 Hongo, Bunkyo-ku, Tokyo 113-0033, Japan}
\address[myaddress2]{The Institute for Solid State Physics, the University of Tokyo, 5-1-5 Kashiwanoha, Kashiwa-shi, Chiba 277-8581, Japan}
\address[myaddress3]{Institute for Materials Research, Tohoku University, 2-1-1 Katahira, Aoba-ku, Sendai 980-8577, Japan}

\begin{abstract}
We have developed a repeating pulsed magnet system which generates magnetic fields of about 10~T in a direction transverse to an incident beam over a length of 0.8~m with a repetition rate of 0.2~Hz.
Its repetition rate is by two orders of magnitude higher than usual pulsed magnets.
It is composed of four low resistance racetrack coils and a 30~kJ transportable capacitor bank as a power supply.
The system aims at axion-like particle searches with a pulsed light source and vacuum birefringence measurements.
We report on the details of the system and its performances.
\end{abstract}


\end{frontmatter}

\linenumbers

\section{Introduction}
A strong magnetic field is one of the most important techniques to study fundamental physics.
In particle physics, it is interpreted as a source of virtual photons and used in many experiments searching for new particles which interact with photons, such as axion-like particles (ALPs).
Since ALPs couple to two photons, experimental searches for ALPs prepare strong magnetic fields over a long distance and convert real photons to ALPs and vice versa.
The conversion probability strongly depends on the integral of the transverse magnetic field $B(z)$ over the optical path length $L$ along the beam axis $z$, and is given by
\begin{equation}
  P(a\leftrightarrow\gamma) = \frac{g_{a\gamma\gamma}^2}{4}\frac{\omega}{\sqrt{\omega^2-m_{a}^2}}\left|\int_{0}^{L}B(z)e^{iqz}{\rm d}z\right|^2,
\end{equation}
where $g_{a\gamma\gamma}$ is the two-photon coupling of ALPs, $\omega$ is the photon energy, $m_{a}$ is the mass of the ALP, and $q$ is the momentum transfer to the magnetic field.
So far many experimental searches for ALPs have been performed with various magnets, such as superconducting dipole magnets~\cite{ALPS, ESRF, OSQAR, SUMICO, CAST} and pulsed magnets~\cite{REVIEW, CNRS}.
In a previous experiment with pulsed magnets~\cite{CNRS}, the magnetic field reaches 12.3~T and is the strongest among previous ALP searches.
However, the repetition rate is limited by Joule heat of the magnet and very low (5 pulses per hour).
Our new repeating pulsed magnet system is designed to resolve this problem.
It especially enables laboratory searches for ALPs in the $\sim$0.1~eV mass region with the highest sensitivity by combining it with a pulsed light source like an x-ray free electron laser (XFEL).

Our magnet system also aims to detect vacuum magnetic birefringence (VMB).
It is one of the fundamental predictions of quantum electrodynamics (QED) that a strong magnetic field transforms an isotropic vacuum into a birefringent medium.
The phase shift of a probe light after passing through the birefringent vacuum can be written as
\begin{equation}
  \psi = \pi k_{{\rm CM}} \frac{B^2 L}{\lambda} \sin 2\theta ,
\end{equation}
where $k_{{\rm CM}}$ ($\sim 4\times10^{-24}$~T$^{-2}$) is the parameter for the linear magnetic birefringence of vacuum, $B$ is the field strength, $L$ is the path length of the probe light in the magnetic field, $\lambda$ is the light wavelength, and $\theta$ is the angle between the polarization of the probe light and the direction of the magnetic field.
The $B^2 L$ dependence requires strong and long transverse magnetic fields to detect vacuum birefringence.
In addition, time variation of the magnetic fields is essential to distinguish VMB signals from static birefringence caused by mirrors, residual gases, residual magnetic fields, and so on.
Fast time variation of pulsed magnetic fields ($\sim$kHz) is of advantage to suppress these static noises, compared to other experiments using superconducting magnets or permanent magnets~\cite{BFRT, PVLAS1, PVLAS2}.
In 2014, the BMV collaboration published a paper on a vacuum birefringence experiment using a pulsed magnet which can generate a maximum field up to 14~T over a length of 0.137~m~\cite{BMV}.
However, the repetition rate of the field generation is 6 pulses per hour, though the experimental sensitivity increases as the square root of the repetition rate.
We have designed our new magnet system so as to increase its repetition rate up to sub-Hz, of course, with higher $B^2 L$.

The repeating pulsed magnet system consists of four racetrack coils and a transportable capacitor bank powering the magnets.
A solenoid magnet for repeating pulsed field generation was first reported in \cite{NOJIRI}, but no racetrack type magnet with a high repetition rate was developed so far.
We have developed low resistance coils to reduce Joule heat during magnetic field generation, and carefully designed its reinforcement to obtain high cooling efficiently of the coils.
The capacitor bank is tailor-made to resupply its capacitor with the energy lost in the coils as Joule heat and enable us to generates high magnetic fields with a high repetition rate.
In this paper, we describe the details of out magnet system and its performances.

\section{Pulsed Magnets}

\begin{figure}[hbtp]
  \centering
  \includegraphics[width=7cm,clip]{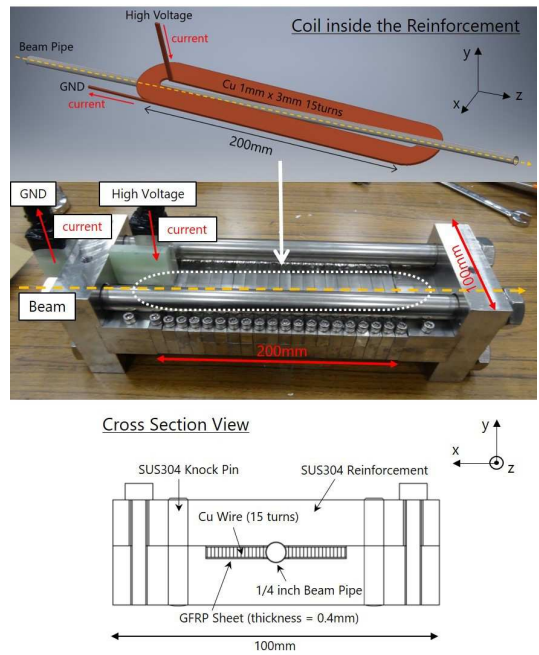}
  \caption{A picture of the magnet, a schematic view of the coil inside the reinforcement made of stainless steel, and the central cross section view of the magnet.}
  \label{fig_coil}
\end{figure}

A schematic view of our coil are shown in Fig.~\ref{fig_coil}.
The coil is composed of 15 turns of a 1$\times$3~mm$^2$ rectangular copper wire.
The length of one racetrack coil is 0.2~m, and a 1/4 inch beam pipe goes through the coil at an angle of 2.75$^{\circ}$.
The coil is impregnated with epoxy resin (Stycast 1266) mixed with alumina powder and reinforced with stainless steel to increase the regidity of the magnet as shown in the picture in Fig.~\ref{fig_coil}.
The reinforcement is divided into electrically insulated 22 slices in a longitudinal direction, in order to reduce an eddy current.

The repetition rate of field generation is limited by Joule heat of the coil and its cooling efficiency.
The magnet is cooled with liquid nitrogen to reduce the coil resistance, because the resistivity of copper at 77~K is nine times smaller than that at 300~K.
In addition, we reduce the thickness of the G10 insulating sheet between the coil and its backup reinforcement as thin as possible (0.4~mm, see the central cross section view of the magnet in Fig.~\ref{fig_coil}), in order to decrease the thermal resistance between the copper wire and the liquid nitrogen.
The calculated thermal resistance is 0.2~K/W.
This means that the thermal time constant of our magnet is expected to be much shorter ($\sim10$~sec) than that of usual pulsed magnets ($1\sim10$~min), considering that the heat capacity of the copper wire of our magnet is about 70~J/K.

\begin{figure}[hbtp]
  \centering
  \includegraphics[width=7cm,clip]{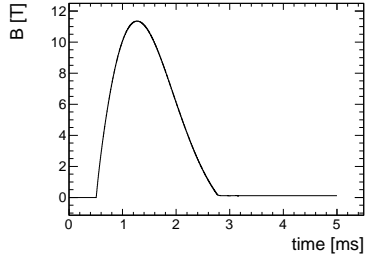}
  \caption{A typical magnetic field pulse of our magnet. A 10~mF capacitor bank was used, and the magnetic field was measured at the center of the beam pipe of the magnet.}
  \label{waveform}
\end{figure}

Let us show the properties and performances of the magnet.
A typical magnetic field pulse generated by our magnet with a 10~mF capacitor bank is represented in Fig.~\ref{waveform}.
The magnetic field was measured at the center of the beam pipe of the magnet.
The pulse width is given by $T=\pi\sqrt{LC}$, where $L$ is the coil inductance and $C$ is the capacitance of a capacitor bank.
The coil inductance of our magnet is $L\sim40$~$\mu$H and the DC resistance is $R=9$~m$\Omega$.
We suppress the increase of the coil resistance at high frequency due to the skin effect or eddy current by dividing the stainless steel for reinforcement, and as a result, the AC resistance is kept sufficiently small (19~m$\Omega$) at 500~Hz.
The combination of the short pulse width and the small resistance decreases Joule heat during pulsed field generation, which is essential for the repeating pulsed field generation with a high repetition rate.

\begin{figure}[hbtp]
  \centering
  \includegraphics[width=7cm,clip]{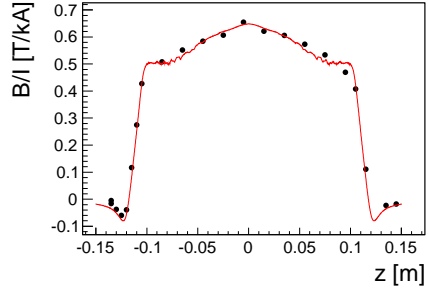}
  \caption{The magnetic field profile of our magnet over the beam axis. The magnetic field is divided by the coil current. The points show the measured profile, while the curve represents the result of finite element simulations. We use four magnets to obtain high magnetic fields over a length of 0.8~m.}
  \label{map}
\end{figure}

\begin{figure}[hbtp]
  \centering
  \includegraphics[width=7cm,clip]{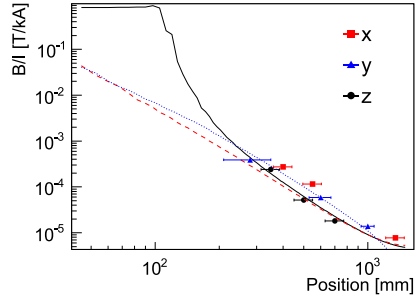}
  \caption{The leakage field of the coil. The circles show the measured leakage field as a function of the distance from the coil in the z-direction which is defined in Fig.~\ref{fig_coil}, while the solid line shows the calculation. The measured and calculated leakage field in the x-direction are shown with the squares and dashed line, respectively. The blue triangles and dotted line correspond to the y-direction.}
  \label{leak}
\end{figure}

The magnetic field profile is shown in Fig.~\ref{map}.
The points represent the measured profile divided by the coil current, while the curve shows the result of finite element simulations with ANSYS~\cite{ANSYS}.
The magnetic field varies along the beam axis because the beam pipe is slightly leaning, and it vanishes rapidly at the both ends of the coil.
The residual magnetic field at a distance of 1~m from the coil center is measured to be of a order of $10^{-4}$~T even when the magnetic field at the center of the coil reaches 10~T as shown in Fig.~\ref{leak}.

\begin{figure}[hbtp]
  \centering
  \includegraphics[width=7cm,clip]{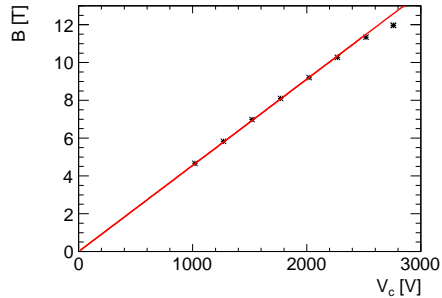}
  \caption{Peak fields at the center of the magnet as a function of the charge voltage. A 10~mF capacitor bank was used during this measurement. The line indicates the result of the linear fitting up to 2500~V.}
  \label{BvsV}
\end{figure}

The most important parameter of the magnet is its maximum field.
It is limited by the mechanical strength of the coil in the case of a pulsed magnet.
Figure~\ref{BvsV} shows peak fields at the center of the magnet as a function of the charge voltage.
A 10~mF capacitor bank was used during this measurement.
The magnet is destructed by a mechanical stress due to the Lorentz force during a pulsed field generation of 12~T.
In order to repeat pulsed field generation, the magnet must be operated below the maximum field, because damages due to the mechanical stresses and heat cycles accumulate.
So far we have generated more than 10,000 magnetic fields of about 9~T, whose mechanical stress is 1.8 times smaller than that of the maximum field, and the coil parameters and the shape of the magnetic field pulses have not changed at all.

\begin{figure}[hbtp]
  \centering
  \includegraphics[width=7cm,clip]{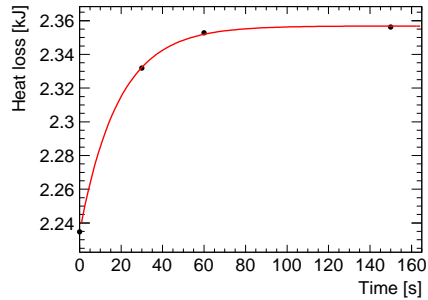}
  \caption{Heat loss of the magnet as a function of the time from the start of repeating pulse generation. The bank energy during the measurement was 3~kJ, and magnetic field pulses of 6.5~T and 4.4~T were generated alternately at 0.4Hz with one magnet. The thermal time constant is obtained to be 19~sec from the fitting.}
  \label{LossVsTime}
\end{figure}

We also measured the cooling efficiency of the magnet which limits its repetition rate.
Since it is difficult to measure the coil temperature directly, we measured the heat loss of the magnet during pulsed field generation, which is proportional to the coil resistance.
The heat loss can be estimated from the decrease of the energy stored in a capacitor bank.
Figure~\ref{LossVsTime} represents the heat loss of the magnet as a function of the time from the start of repeating pulsed field generation.
The bank energy during the measurement was 3~kJ, and magnetic field pulses of 6.5~T and 4.4~T were generated alternately at 0.4Hz with one magnet.
The thermal time constant is estimated to be 19~sec from the fitting.
We also measured time constants with various conditions of charge voltages, repetition rates, and the number of magnets, and the estimated time constants are between 13~sec and 26~sec.
These time constants are much shorter compared to those of usual pulsed magnets ($1\sim10$~min), and enable repeating pulse generation with a high repetition rate.

\section{Capacitor Bank}

\begin{figure}[hbtp]
  \centering
  \includegraphics[width=12cm,clip]{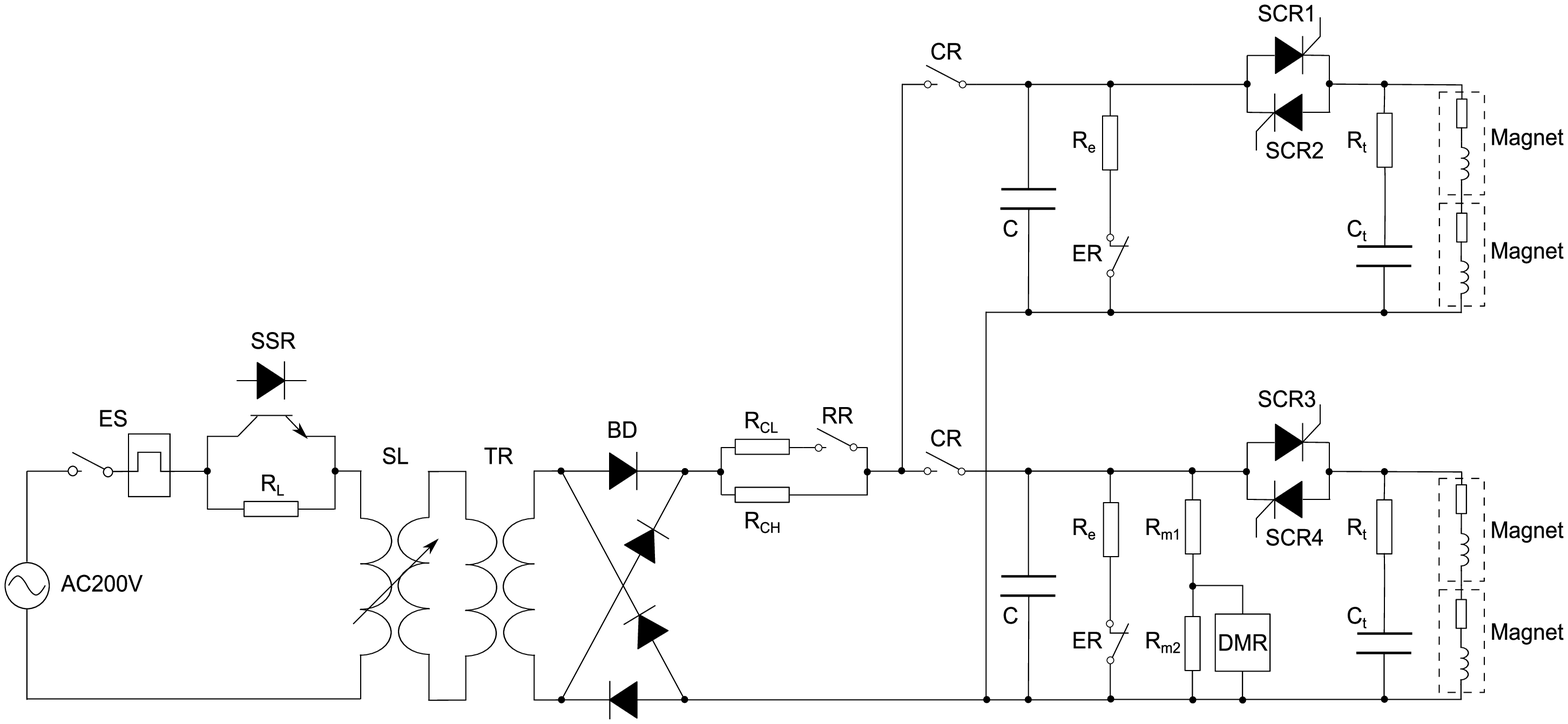}
  \caption{The circuit diagram of the capacitor bank. ES : electrical switch, SSR : solid state relay, SL : variable transformer, TR : transformer, BD : diode bridge rectifier, DMR : digital meter relay, SCR : thyristor.}
  \label{circuit} \end{figure}

\begin{figure}[hbtp]
  \centering
  \includegraphics[width=7cm,clip]{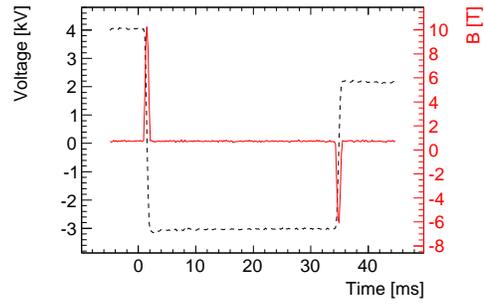}
  \caption{Typical waveforms during a repeating operation with an external trigger of 30~Hz. The dashed line shows the voltage of the capacitors, and the magnetic field pulse is shown in the solid line.}
  \label{g_VandI}
\end{figure}

We have also developed a transportable capacitor bank to supply a large electric current ($\sim$30~kA) to the coils.
Its circuit diagram is shown in Fig.~\ref{circuit}.
We usually connect four magnet coils (total length~$=0.8$~m) to the capacitor bank.
The capacitor bank has two 1.5~mF capacitor stacks (C in Fig.~\ref{circuit}) and they can be charged up to 4.5~kV (maximum energy~$=$~30~kJ).
One capacitor stack consists of six 250~$\mu$F capacitors (Nichicon CCFI-652450HGW).
We use thyristors as discharge switches, and place two thyristor stacks in antiparallel (SCR1\&SCR2, SCR3\&SCR4 in Fig.~\ref{circuit}) to select the direction of the electric current through the coils.
One thyristor stack is composed of four thyristors (Dean Technology, T77P3000S12100) connected in series to increase its breakdown voltage.
Figure~\ref{g_VandI} shows typical waveforms of a magnetic field and capacitor voltage during a repeating operation triggered by an external clock of 30~Hz.
In order to increase the repetition rate of pulsed field generation, we operate the capacitor bank as follows.
First, SCR1 and SCR3 in Fig.~\ref{circuit} are switched on at the timing of the external clock and the energy charged in the capacitors goes through the coils as an electric current.
Some of the energy is consumed in the coil as Joule heat, and the rest of the energy charges the capacitors again with reversed polarity.
Next, SCR2 and SCR4 are turned on and the reversal process occurs in synchronization with the external clock.
Finally, we recharge the capacitor bank with a charging circuit without accepting external triggers.
Since the Joule heat of the coils is very small, the recharging process takes much shorter time than initial charging from 0~V.

\begin{figure}[hbtp]
  \centering
  \includegraphics[width=5cm,clip]{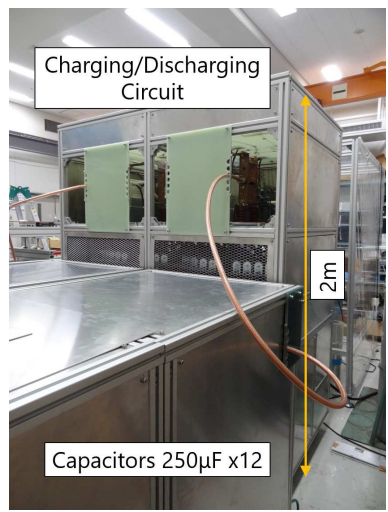}
  \caption{A picture of the 30kJ transportable capacitor bank to generate repeating pulsed magnetic fields.}
  \label{pic_bank}
\end{figure}

Figure~\ref{pic_bank} is a picture of the capacitor bank.
It occupies a space of about $2\times3\times2$~m$^3$ (Width$\times$Depth$\times$Height) and weighs about 2.5~tons, which is small enough to fit in an experimental hatch of a synchrotron radiation facility.
It can be easily divided into six parts so that we can deliver it to other facilities such as SACLA (SPring-8 Angstrom Compact Free Electron Laser)~\cite{SACLA}, which is one of XFEL sources, to perform experiments.

\begin{figure}[hbtp]
  \centering
  \includegraphics[width=12cm,clip]{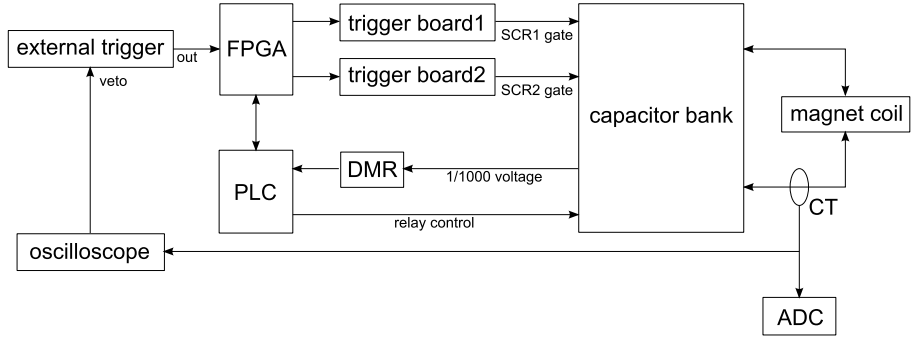}
  \caption{The Block diagram of the control system.}
  \label{blockdiagram}
\end{figure}

The control system of the capacitor bank consists of a triggering system to generate a magnetic field at the timing of an external trigger, a sequencer to control many charge and discharge relays, and a waveform checker for safety.
Its block diagram is represented in Fig.~\ref{blockdiagram}.
The charge voltage of the capacitor bank is divided to 1/1,000 and the divided voltage is monitored with a digital meter relay using output photocouplers to achieve a high speed operation up to $\sim$kHz (DMR, Watanabe Electric Industry Co. Ltd., AMH-148-VA-12).
The DMR and a programmable logic controller (PLC) work together to control high voltage relays in the capacitor bank sequentially.
We use a FPGA board and two gate drive circuit boards to drive thyristors and generate magnetic field pulses at the timing of the external trigger.
It enables us to synchronize the field generation with a pulsed light in the search for ALPs.

The current of the coils is monitored with a current transformer (CT, Pearson Electronics, Model 1423).
All waveforms of the pulsed field generation are recorded with an analog-to-digital converter (ADC).
In addition, all waveforms are tested on an oscilloscope (Rohde \& Schwarz, RTO1044) by comparing them with a template waveform saved before a repeating operation.
If an irregular waveform is found, the oscilloscope immediately send a trigger output to veto the external trigger and stop pulse generation.

\begin{figure}[hbtp]
  \centering
  \includegraphics[width=7cm,clip]{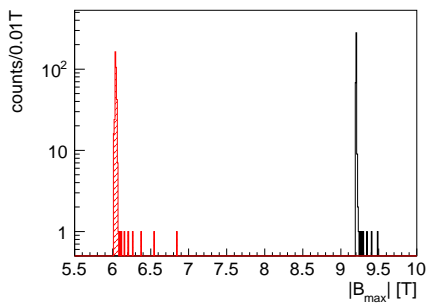}
  \caption{The result of the repeating pulsed field generation during an operation for one hour. The solid line and hatched histogram shows the absolute values of maximum magnetic fields of first and second pulses, respectively.}
  \label{waveform_accum}
\end{figure}

\begin{figure}[hbtp]
  \centering
  \includegraphics[width=7cm,clip]{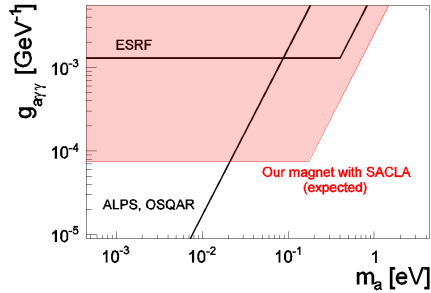}
  \caption{Expected sensitivity of the ALP search by combinining our repating pulsed magnet with an XFEL is shown in the filled region. We also show the limits for the ALP-photon coupling obtained from previous laboratory searches using superconducting magnets with a synchrotron radiation source (ESRF~\cite{ESRF}) and with infrared or optical laser sources (ALPS~\cite{ALPS}, OSQAR~\cite{OSQAR}) in the solid lines.}
  \label{reach}
\end{figure}

The result of the repeating pulsed field generation with our four magnet coils and our transportable capacitor bank is shown in Fig.~\ref{waveform_accum}.
The histogram shows the absolute values of maximum magnetic fields of first pulses generated with 4~kV discharge and the hatched histogram shows those of second pulses with 3~kV discharge.
Magnetic field pulses of 9.2~T and 6.0~T over a length of 0.8~m were generated alternately, and the total number of pulses were 732 during an operation for one hour.
The repetition rate is about 0.2~Hz.
Though the peak magnetic fields gradually decrease for about 100 seconds from the start of an operation because the temperature of the coils increases with their thermal time constants ($\sim20$~sec), the variation of magnetic fields during a steady operation is less than $\pm0.5$\%.
The average values of $B^2L$ (for the VMB experiment) and $BL$ (for the ALP search) of first pulses are 54~T$^2$m and 7.1~Tm, respectively.
In addition, more than 10,000 magnetic fields of about 9~T have been generated without any damages to the coils so far.
In Fig.~\ref{reach}, we show the expected sensitivity of the ALP search with our magnet system and an XFEL.
High intensity x-ray pulses from an XFEL is well suited to be combined with repeating pulsed magnetic fields.
In addition, the short ($<10$~fs) XFEL pulse enables a background-free experiment.
Assuming a hard ($\sim10$~keV) x-ray pulse comprises $5\times10^{11}$ photons and two days of data taking, we can expect the highest sensitivity to the ALP-photon coupling $g_{{\rm a}\gamma\gamma}$ in the high-mass region $m_{\rm a}\sim0.1$~eV.

\section{Conclusion}
We have developed a repeating pulsed magnet system to generate transverse magnetic fields of about 10~T over a length of 0.8~m with a repetition rate of 0.2~Hz.
The repetition rate is higher than usual pulsed magnets by two orders of magnitude.
It consists of four low resistance racetrack coils and a 30~kJ transportable capacitor bank optimized to repeat pulsed fields.

The combination of the repeating pulsed magnets with pulsed light sources enables us to search for ALPs with high sensitivity.
For instance, we can search for relatively massive ALPs in the $\sim$0.1~eV mass region with the highest sensitivity by using an XFEL, whose intense pulsed beam with a high repetition rate matches our magnet system.
The pulsed magnet system is also best suited to measure vacuum magnetic birefringence because the time variation of the magnetic field is essential to detect VMB signals among a lot of static matter birefringence.
Its sensitivity to the birefringence of vacuum is 100 times higher than the previous measurement using a pulsed magnet.

We plan further improvements of the coils and the capacitor bank.
In order to increase magnetic fields up to over 20~T, we are investigating a multi-layer coil made of Cu-Ag alloy whose tensile strength is more than three times higher than that of copper wire.
However, the resistivity of Cu-Ag alloy is also higher than that of copper.
This means more Joule heat, therefore we have to design appropriate reinforcement and insulating structure which keep good cooling efficiency with sufficient mechanical strength.
An upgrade of the capacitor bank will also be performed by adding capacitors because the maximum $B^2L$ of the magnet system is limited by the energy stored in the capacitor bank.

\section*{Acknowledgements}
This work was performed under the Inter-university Cooperative Research Program of the Institute for Materials Research, Tohoku University (Proposal No.~14K0018 and 15K0080), and supported by JSPS KAKENHI Grant Number 13J07172 and MEXT KAKENHI Grant Number 23224009 and 26104701.


\bibliography{mybibfile}

\end{document}